\documentclass[journal]{IEEEtran}
\usepackage[T1]{fontenc}
\usepackage[utf8]{inputenc}
\usepackage{amsmath,amssymb,array,booktabs,graphicx,textcomp,microtype}
\usepackage[hidelinks,pdftitle={Clean Accuracy Does Not Guarantee Provenance Robustness: A Prospective Codec-Stress Evaluation of Audio Attribution},pdfauthor={Gang Shi}]{hyperref}
\begin{document}
\title{Clean Accuracy Does Not Guarantee Provenance Robustness: A Prospective Codec-Stress Evaluation of Audio Attribution}
\author{Gang~Shi%
\thanks{G. Shi is an independent researcher (e-mail: rockyshicoder@gmail.com). This work was conducted independently and without institutional support.}}
\maketitle

\begin{abstract}

Audio provenance attribution --- which system produced a synthetic utterance --- is reported at near-ceiling accuracy on clean benchmarks, yet audio reaching an analyst has usually been transcoded. We report a prospectively registered measurement of closed-set attribution after single-stage codec transport, with the analysis region fixed from fidelity metadata before any attribution model was trained. On two corpora, in-support losses reach 53.5 [43.5, 63.6] and 70.3 [63.0, 77.5] Macro-F1 points for WavLM-Base+, and 61.0 [56.8, 65.1] and 49.8 [41.6, 57.9] for W2V2-BERT 2.0, under simultaneous component-level bands. Degradation is strongly condition- and representation-dependent: within one in-support grid WavLM losses run from $-$0.4 to +53.5 points, and the two encoders differ beyond a prespecified $\pm$5-point margin at six of twelve conditions. A clean-qualified ECAPA-TDNN and a Proxy-Anchor head degrade comparably, so the effect is not confined to one representation family or a weak linear head. The registered matched-fidelity comparison was not estimable on this grid, and waveform and perceptual measures order the conditions differently: MP3 at 8 kbit/s ranks mid-grid on SI-SDR but last on PESQ-WB while causing the largest loss. For the tested tasks, corpora, representations and codec grid, a clean accuracy figure does not by itself characterise deployment robustness.

\end{abstract}

\section{Introduction}\label{sec:I}

Attribution of synthetic speech to its generating system is approaching saturation on clean benchmarks \cite{ref2}. Forensic use is almost never clean-set use: audio reaching an analyst has passed through telephony, messaging, social platforms or archival storage, compressed with MP3, Opus, or increasingly a neural codec, sometimes more than once.

The question is not whether lossy compression degrades a classifier --- it plainly does. It is a measurement question an engineer has to answer before deploying such a system: \textbf{how much of a reported clean attribution score survives realistic transport, and can that be predicted from the clean score alone?}

Answering it credibly is harder than running codecs over a test set. Codec conditions differ on several distortion axes at once, so a family comparison confounds family with severity, and both the grid and the analysis can be adjusted after seeing outcomes. Section~\ref{sec:III} gives the evaluation boundary we fixed in advance.

This paper makes three contributions, each bounded to the tested closed-set provenance tasks, two corpora, three representations and single-stage codec grid.

\textbf{C1.} Clean attribution accuracy does not guarantee robustness after codec transport. All three tested representations exceed 0.85 clean Macro-F1 and two exceed 0.97, yet lose up to 53.5 and 70.3 Macro-F1 points at a single in-grid operating point.

\textbf{C2.} Codec-stress degradation is strongly condition- and representation-dependent. Within the same in-support grid, losses that are indistinguishable from zero coexist with losses above 50 points, and the two deciding representations differ beyond a prespecified $\pm$5-point margin at six of twelve conditions on one corpus, with the ordering between them reversing across corpora. One clean score, or one representation, is therefore insufficient to characterise deployment robustness.

\textbf{C3.} The preregistered matched-fidelity comparison lacked sufficient common support on the tested codec grid, and the waveform and perceptual measures rank the codec conditions differently --- MP3 at 8 kbit/s sits mid-grid on SI-SDR and last on PESQ-WB on both corpora. A single fidelity scalar therefore does not order codec severity for this purpose.

We propose no mitigation. The aim is to bound what these systems can be trusted to do once the audio has been through a codec.

\section{Related Work}\label{sec:II}

\textbf{Clean and open-set source-model attribution.} Attribution has moved from binary spoof detection to multi-class identification of the generating system. M\"uller et al.'s MLAAD \cite{ref1} supplies the substrate most recent work shares, and on it Neamtu et al. \cite{ref2} report 99.76 \% closed-set accuracy over 110 TTS architecture classes with FPR@95 as low as 2.04 \%, using Proxy-Anchor metric learning over frozen W2V2-BERT embeddings. Klein et al. \cite{ref3} address the open-set variant on the same corpus family; Xie et al. \cite{ref4} show that models strong on in-distribution classification and OOD rejection still fail on unseen \textit{real} audio; and Firc et al. \cite{ref5} and Rubio et al. \cite{ref6} reject the multiclass-plus-fallback framing altogether, the first for a NIST-SRE-style trial protocol and the second for a compositional view of what a source is. Pizarro et al. \cite{ref7} show a training-free residual statistic reaches AUROC above 99 \% for open-world single-model attribution. Because these three open-set framings \cite{ref3,ref5,ref6} are mutually incompatible, we adopt a closed-set multiclass task throughout and make no open-set claim. We take the Proxy-Anchor objective of \cite{ref2} as our strong-method comparator, in a deliberately reduced-scale instantiation, so that a failure cannot be attributed to a weak linear boundary alone.

\textbf{Codec and channel robustness in deepfake detection.} The detection literature has measured transformation robustness; our literature audit found no comparable measurement for attribution. Li et al. \cite{ref8} run 18 corruptions across 10 detectors and find noise survivable while modification and compression are not; Gao et al. \cite{ref9} report 22 detectors losing 43 \% under realistic conditions; Li et al. \cite{ref10} identify platform re-encoding as a dominant real-world failure mode; and Cohen et al. \cite{ref11} established early that compression and channel augmentation recovers much of the loss for detection. This body of work supplies our detection-side analogue but not our answer: these are binary detection results, and \cite{ref8} is a single-corpus sweep over six vocoders from one speaker. Whether a \textit{multi-class attribution} decision survives transport, and how its uncertainty behaves on grouped sources, is not addressed there. Our literature audit found no source-tracing work reporting attribution performance as a function of codec family and operating point; we state this as the audit's position rather than as an absence claim about the field.

\textbf{Source tracing and neural-codec attribution.} Neural codecs are simultaneously a transport mechanism and a forensic signal, which is the confound at the centre of this paper. Moussa et al. \cite{ref12} show that neural codecs leave distinctive frequency artefacts that identify the compression codec alone, with no generator involved, establishing that codec identity is itself decodable. Chen et al. \cite{ref13} localise the mechanism, showing that disentanglement objectives and frequency-domain decoders drive codec artifacts, and Chen et al. \cite{ref14} decompose codec-based generation for tracing. Phukan et al. \cite{ref15} go further and regress codec \textit{parameters} rather than predicting a label. Against this, Cuccovillo et al. \cite{ref16} argue that betting synthetic-speech detection solely on neural-encoding artifacts is an over-specialisation; Section~\ref{sec:IX} bears on that debate. Our question is oriented differently from all of this work. There the codec is the provenance signal itself, or the training substrate used to expose it; here a provenance task is already defined --- by TTS system or by codec source --- and we ask whether a \textit{subsequent transport codec} disrupts the attribution decision that task supports. The codecs themselves are EnCodec \cite{ref17} and DAC \cite{ref18}.

\textbf{Robustness diagnosis, shortcuts, and codec cascades.} A diagnostic literature exists for detection and has repeatedly found that apparent success rests on artifacts rather than the phenomenon: dataset artefacts in ASVspoof 2017 \cite{ref19}, silence statistics \cite{ref20}, and speaker identity as a shortcut \cite{ref21}. Our literature audit found no application of that methodology to attribution, which is part of why we register support boundaries and disjointness gates rather than assuming them. On the codec side, Tseng and Harwath \cite{ref22} probe neural-codec robustness directly and find non-linear distortion partly explains differences between codecs, and Park et al. \cite{ref23} show that for neural-codec token statistics the acoustic condition dominates corpus identity; both are consistent with Section~\ref{sec:VII}. Cascade forensics exists for traditional compression, where Xiang et al. \cite{ref24} localise single versus multiple MP3 compression at frame level, but we are not aware of a neural-codec analogue; our grid includes single-stage conditions only, and cascades remain future work.

\section{Evaluation Design and Evidence Discipline}\label{sec:III}
\begin{table*}[t]
\caption{Design and effective inference units. Clip counts are large; the number of independent units is not, and every interval is computed on components.}
\label{tab:4}
\centering\small
{\hyphenpenalty=9000\exhyphenpenalty=9000
\begin{tabular}{@{}>{\raggedright\arraybackslash}p{0.312\linewidth}>{\raggedright\arraybackslash}p{0.612\linewidth}@{}}
\toprule
quantity & value \\
\midrule
independent components, LA & 45 \\
independent components, ST-Codecfake-OOD & 28 \\
bootstrap & B = 1000, multiplicity-preserving; stratification varies by analysis (Section~\ref{sec:III}) \\
simultaneous bands & max-$|t|$ over the registered family \\
GPU wall-clock & 9.34 h against an 18 h registered ceiling \\
transformation operations & 57,057, zero failures \\
\bottomrule
\end{tabular}}
\end{table*}

This section states the facts needed to read the results. Process detail beyond that will be released with the evaluation artifacts.

\textbf{Development and prospective evidence are kept separate.} Hypotheses were generated on development data and tested on separate prospective data under a protocol fixed in advance. The two are never pooled, and only prospective evidence supports Sections V--IX.

\textbf{The registered matched-fidelity comparison was not estimable on this grid.} The original primary comparison paired conventional and neural codec conditions matched simultaneously on |$\Delta$SI-SDR| $\leq$ 2.0 dB, |$\Delta$PESQ-WB| $\leq$ 0.35, |$\Delta$STOI| $\leq$ 0.05 and |$\Delta$bandwidth| $\leq$ 1.0 kHz, one-to-one without replacement, gated at a minimum of two admissible pairs per corpus. Evaluated on calibration fidelity data alone --- before any attribution embedding existed --- the rule admitted exactly one pair per corpus and failed the gate on both. The comparison is recorded as \textbf{not estimable under the preregistered matching rule and tested grid} (Section~\ref{sec:VII}), a verdict scoped to this rule, these tolerances, this grid and these corpora. It is not a claim that matched-fidelity codec comparison is impossible in general: a wider grid, looser tolerances, or codecs chosen to overlap in fidelity could restore support, at the cost of a different estimand.

\textbf{A conditional-association analysis replaced it, and is secondary.} Because matching was unavailable we registered an amendment (Q1b) on a common-support region --- a box on SI-SDR $\times$ STOI formed by intersecting the 5th--95th percentile ranges of the two families, admitting only conditions contributing at least 20 \% of their observations. That support rule was selected on calibration data after a draft rule failed there, so it is calibration-informed and post-registration. We therefore report Q1b as a secondary exploratory analysis (Section~\ref{sec:VIII}), not as a headline result.

\textbf{The support boundary was frozen before any attribution outcome.} The mask was computed from fidelity metadata only, then materialized, hashed and committed \textbf{before any attribution head was trained and before any attribution outcome was computed or inspected}. ASVspoof 2019 LA admitted a support box of 5.8--15.4 dB SI-SDR $\times$ 0.866--0.991 STOI, with 5 contributing conventional and 7 contributing neural conditions; ST-Codecfake-OOD admitted 8.7--15.9 dB $\times$ 0.835--0.986 with 4 and 5. Both passed their registered admission thresholds. A third corpus, MLAAD-unseen, produced an \textbf{empty} support box and was designated \textit{sensitivity only} in advance; that designation was not repaired after the fact.

\textbf{Independence and units.} The unit of inference is the \textit{independent component}, defined operationally as follows. Build a graph whose nodes are clips, and join two clips whenever they share a registered source identifier or a registered speaker identifier; an independent component is a connected component of that graph. On ASVspoof 2019 LA the source identifier is the synthesis prompt text and the speaker identifier is the target speaker, so every clip generated from one prompt, and every clip sharing a target speaker, is drawn into the same component. On ST-Codecfake-OOD the source identifier is the originating natural utterance and the speaker identifier is that utterance's speaker. Data admission is fail-closed with pre-declared branches only, and encoders enter the stress analysis only after passing a clean qualification gate (Table~\ref{tab:1}).

\textbf{Splits, class hold-out, and what the five seeds are.} Each seeded run first withholds 25 \% of the \textit{provenance classes} entirely --- they appear in no partition --- and the remaining in-distribution classes are then allocated over whole components in a 60/10/30 train/validation/test partition, so no source and no speaker crosses a split boundary. A class is retained for scoring only if at least 20 of its clips land in that seed's test partition. The registered analysis runs five seeds.

Those five seeded runs do \textbf{not} realise five distinct splits. The hold-out is drawn by shuffling the class index under the seed, and the component allocation that follows is a deterministic greedy rule, so two seeds drawing the same hold-out set produce an identical split --- identical down to the stored per-clip predictions. On the small label spaces used here (7 classes on LA with 2 withheld, 5 classes on ST-Codecfake-OOD with 1 withheld) repeat draws are likely, and they occurred: the five seeded runs correspond to \textbf{three unique split configurations} on each corpus, at multiplicities 2/1/2 on LA (seeds {0,1}, {2}, {3,4}) and 3/1/1 on ST-Codecfake-OOD (seeds {0,1,2}, {3}, {4}). Duplicate configurations therefore carry repeated weight in the registered five-seed aggregate.

The registered five-seed analysis remains the primary analysis throughout this paper. A post-hoc sensitivity analysis that instead weights the three unique configurations equally is reported wherever it changes a qualification decision; it changes no primary decision (Sections V, VI, VIII and XI).

This leaves \textbf{45 independent components on LA and 28 on ST-Codecfake-OOD} (Table~\ref{tab:4}), against clip counts an order of magnitude larger. \textbf{Clip count is not the inferential sample size.} Every interval reported in this paper is computed on components.

\textbf{Bootstrap.} Uncertainty is a cluster bootstrap that resamples components rather than clips, with B = 1000 replicates; the single exception is Q2, whose cluster is the source utterance (Section~\ref{sec:IX}). A cluster drawn k times contributes all of its rows k times, so resample multiplicity is preserved rather than collapsed, one draw is shared across encoders, conditions and split seeds, and simultaneous bands are max-|t| bands over the registered family of quantities. \textit{Class-signature stratification}, where used, means: a component's signature is the set of provenance classes its clips carry, components sharing a signature form one stratum, and each stratum is resampled to its own size, so a replicate cannot alter the class composition of the sample.

Stratification and interval construction are \textbf{not} uniform across analyses:

\begin{center}\small
{\hyphenpenalty=9000\exhyphenpenalty=9000
\begin{tabular}{@{}>{\raggedright\arraybackslash}p{0.306\linewidth}>{\raggedright\arraybackslash}p{0.189\linewidth}>{\raggedright\arraybackslash}p{0.160\linewidth}>{\raggedright\arraybackslash}p{0.233\linewidth}@{}}
\toprule
analysis & unit & stratification & interval \\
\midrule
Stress, interaction, PA, generality (V--VI) & component, per corpus & class signature & simultaneous \\
Non-codec controls (VII) & component, pooled & class signature & simultaneous \\
Q1b registered primary (VIII) & component, pooled & \textbf{none} & simultaneous; percentile for $\Delta R^2$ \\
Q1b class-signature audit (sensitivity) & component, pooled & class signature & as primary \\
Q2 overwrite rule (IX) & source cluster, per corpus & source label signature & percentile \\
\bottomrule
\end{tabular}}
\end{center}

All five use B = 1000. In Q2 the real reference bank is resampled separately, by speaker.

The registered Q1b primary bootstrap resamples components \textbf{without} stratification; the class-signature-stratified variant is a sensitivity audit, not the registered primary, and it leaves the H1--H4 truth-table decisions unchanged (Section~\ref{sec:VIII}).

\textbf{Frozen estimands and preserved nulls.} All decision thresholds were fixed before execution: the 20-point simultaneous lower bound and $\pm$5-point encoder-interaction margin for the primary stress decision, $\delta_D$ = 5 points for the association hypotheses, and the conjunct margins of the overwrite rule of Section~\ref{sec:IX}, which was frozen --- as executable code, verified byte-identical to the development implementation --- before any prospective evaluation. Null and negative outcomes are preserved rather than reinterpreted.

\section{Experimental Setup}\label{sec:IV}
\begin{table}[t]
\caption{Clean qualification gates (registered, evaluated before stress). MLAAD-unseen fails every gate and is reported as external-domain sensitivity only.}
\label{tab:1}
\centering\small
{\hyphenpenalty=9000\exhyphenpenalty=9000
\begin{tabular}{@{}llll@{}}
\toprule
corpus & encoder & clean Macro-F1 & gate \\
\midrule
la & ecapa (l=1) & 0.856 & PASS \\
la & w2v2bert (l=3) & 0.971 & PASS \\
la & wavlm (l=4) & 0.982 & PASS \\
st-codecfake-ood & ecapa (l=1) & 0.868 & PASS \\
st-codecfake-ood & w2v2bert (l=3) & 0.967 & PASS \\
st-codecfake-ood & wavlm (l=4) & 0.916 & PASS \\
mlaad-unseen & ecapa (l=2) & 0.664 & FAIL \\
mlaad-unseen & w2v2bert (l=3) & 0.831 & FAIL \\
mlaad-unseen & wavlm (l=4) & 0.745 & FAIL \\
\bottomrule
\end{tabular}}
\end{table}

\textbf{Corpora.} ASVspoof 2019 LA, restricted to a 7-class TTS attack label space (A07--A12 and A16), and ST-Codecfake-OOD, whose 5 classes are untouched codec sources. MLAAD-unseen is a designated external-domain sensitivity arm.

\textbf{The scored task is smaller than the label space, and on LA its size varies by seed.} After the 25 \% class hold-out and the 20-clip minimum test-support rule (Section~\ref{sec:III}), the scored problem on LA is \textbf{3-way} in the two seeded runs that withhold {A09, A16} --- there A07 and A08 additionally fall below minimum test support --- and \textbf{5-way} in the other three, which withhold {A08, A11} and {A11, A16} respectively. On ST-Codecfake-OOD it is \textbf{4-way} in every seeded run, one codec class being withheld each time. All Macro-F1 values are computed over the classes retained in that seed, so a reported LA figure averages runs of differing task size; ST-Codecfake-OOD figures do not. This follows from the registered protocol rather than any post-hoc choice, and it is a further reason we do not compare attribution magnitudes across corpora.

\textbf{Representations.} WavLM-Base+ (hidden state 4) and W2V2-BERT 2.0 (hidden state 3) are the two deciding encoders. ECAPA-TDNN, a non-Transformer speaker-verification representation, is included as a descriptive generality probe. All encoders are frozen; no fine-tuning is performed.

\textbf{Heads.} The linear head is multinomial logistic regression on features standardised with training-split statistics, its inverse-regularisation constant selected on the validation split from the frozen grid $C \in \{0.1, 0.3, 1, 3\}$ by validation Macro-F1. The metric-learning head is a Neamtu-aligned reduced-scale \textbf{Proxy-Anchor} classifier: one linear projection to 1024 dimensions, one learnable proxy per retained class, cosine similarity between $L_2$-normalised embeddings and $L_2$-normalised proxies, Proxy-Anchor loss at scale $\alpha = 32$ and margin $\delta = 0.1$, optimised by AdamW (learning rate $10^{-3}$, weight decay $10^{-4}$, batch size 256) for 100 epochs with the checkpoint selected by best validation Macro-F1. It is trained on W2V2-BERT 2.0 hidden state 4 --- the registration-time metric-learning layer, one layer above the hidden state 3 used by the deciding W2V2-BERT encoder --- and is run at three head seeds per split seed and averaged. It is included so that the failure cannot be attributed to a weak linear boundary alone.

\textbf{Conditions.} MP3, Opus, EnCodec and DAC at multiple operating points, plus four non-codec controls (a 4 kHz bandwidth bottleneck, 8-bit \textmu{}-law, and additive noise at 10 dB and 5 dB SI-SDR). All clips are 16 kHz mono before any transformation; MP3 and Opus are encoded at 16 kHz with ffmpeg 7.0.2, and the neural codecs use a 16 $\rightarrow$ 24 $\rightarrow$ 16 kHz resampling path. Per-clip SI-SDR, PESQ-WB, STOI and bandwidth are measured on latency-aligned audio. MP3 and Opus conditions are named by nominal bit rate in kbit/s, so \texttt{mp3\_8} is MP3 at 8 kbit/s. The neural-codec labels are defined in Table~\ref{tab:50}.

\begin{table*}[t]
\caption{Neural-codec conditions: paper label, implementation, registered configuration and nominal rate. The N2 labels denote DAC; see the text for the pre-registered binding gate.}
\label{tab:50}
\centering\small
{\hyphenpenalty=9000\exhyphenpenalty=9000
\begin{tabular}{@{}>{\raggedright\arraybackslash}p{0.329\linewidth}>{\raggedright\arraybackslash}p{0.131\linewidth}>{\raggedright\arraybackslash}p{0.280\linewidth}>{\raggedright\arraybackslash}p{0.148\linewidth}@{}}
\toprule
paper label & implementation & registered configuration & nominal rate \\
\midrule
encodec\_24k, encodec\_6k, encodec\_3k, encodec\_1p5k & EnCodec \cite{ref17} & 24 kHz model & 24, 6, 3, 1.5 kbit/s \\
N2\_high, N2\_mid, N2\_low & DAC \cite{ref18} & 24 kHz model, $n_q$ = 13, 6, 3 codebooks & 6.1, 2.8, 1.4 kbit/s \\
dac\_nq20 & DAC \cite{ref18} & 24 kHz model, $n_q$ = 20 codebooks & 9.4 kbit/s \\
\bottomrule
\end{tabular}}
\end{table*}

\textbf{The N2 labels denote DAC and are retained for provenance.} \texttt{N2} was the pre-registered slot for the second neural codec, bound by a gate fixed before any attribution outcome existed: bind DAC if its full-codebook decode reached at least 15 dB latency-aligned SI-SDR on speech, otherwise bind Mimi. The gate passed at 16.1 dB, so N2 is DAC throughout and the Mimi branch was never taken. The three \texttt{N2\_*} points are the development-retained DAC operating points and keep their registration-time names; \texttt{dac\_nq20} is the DAC point selected prospectively at registration and is named after the binding was known. Both families are the same codec. DAC rates follow from the 24 kHz model's frame rate of 46.875 s$^{-1}$ at 10 bit per codebook, i.e. 0.469 kbit/s per codebook.

\textbf{Two provenance label spaces.} The two confirmatory corpora instantiate different provenance tasks: on ASVspoof 2019 LA the label is TTS-system identity (7 attack classes, A07--A12 and A16), and on ST-Codecfake-OOD the label is codec-source identity (5 untouched codec classes). We include both to test whether codec-transport degradation reproduces \textit{within} each task definition, not to compare the tasks. We therefore do not compare attribution magnitudes across the two corpora; replication here concerns whether transport codecs induce material within-corpus degradation under each provenance definition.

\textbf{Metric.} True multiclass Macro-F1 on retained predicted classes. Degradation is $\Delta$Macro-F1 = clean -- transformed, in points, so \textbf{positive means loss}.

\section{Result 1: Codec Transport Removes a Large Fraction of Attribution Performance}\label{sec:V}
\begin{figure*}[t]
\centering
\includegraphics[width=\textwidth]{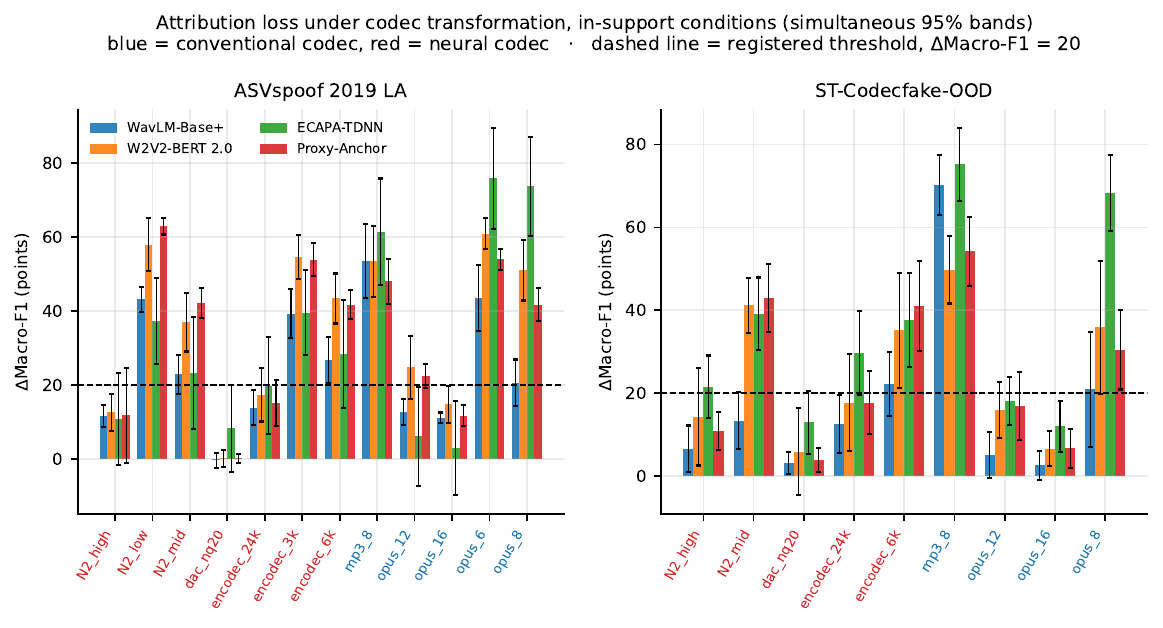}
\caption{Attribution loss under codec transformation at in-support conditions, with simultaneous 95\% bands over components. Blue condition labels are conventional codecs, red are neural. The dashed line marks the registered threshold at $\Delta$Macro-F1 = 20 points; the registered rule requires a condition's simultaneous lower bound, not its point estimate, to lie above it.}
\label{fig:1}
\end{figure*}

All three encoders pass the registered clean qualification gate on both confirmatory corpora, at 0.856 to 0.982 Macro-F1 (Table~\ref{tab:1}). Under the frozen support mask, the largest in-support losses are (Fig.~\ref{fig:1}):

\begin{center}\small
{\hyphenpenalty=9000\exhyphenpenalty=9000
\begin{tabular}{@{}>{\raggedright\arraybackslash}p{0.312\linewidth}>{\raggedright\arraybackslash}p{0.296\linewidth}>{\raggedright\arraybackslash}p{0.297\linewidth}@{}}
\toprule
corpus & model & largest in-support $\Delta$Macro-F1 (points) \\
\midrule
LA & WavLM-Base+ & mp3\_8 \textbf{+53.5} [+43.5, +63.6] \\
LA & W2V2-BERT 2.0 & opus\_6 \textbf{+61.0} [+56.8, +65.1] \\
LA & ECAPA-TDNN (descriptive) & opus\_6 \textbf{+76.0} [+62.3, +89.6] \\
LA & Proxy-Anchor & N2\_low \textbf{+63.0} [+60.7, +65.3] \\
ST-Codecfake-OOD & WavLM-Base+ & mp3\_8 \textbf{+70.3} [+63.0, +77.5] \\
ST-Codecfake-OOD & W2V2-BERT 2.0 & mp3\_8 \textbf{+49.8} [+41.6, +57.9] \\
ST-Codecfake-OOD & ECAPA-TDNN (descriptive) & mp3\_8 \textbf{+75.2} [+66.5, +84.0] \\
ST-Codecfake-OOD & Proxy-Anchor & mp3\_8 \textbf{+54.2} [+45.8, +62.6] \\
\bottomrule
\end{tabular}}
\end{center}

The registered decision required, per corpus, at least one in-support condition per deciding encoder whose simultaneous lower bound exceeds 20 points. WavLM clears it at 5 LA conditions and W2V2-BERT at 7; on ST-Codecfake-OOD, at 1 and 3 respectively. The decision is \textbf{supported on both corpora}.

Two additional arms bound the interpretation. Neither enters the registered decision, and each carries a weighting sensitivity.

ECAPA-TDNN, a speaker-verification representation on a non-Transformer architecture, reaches a simultaneous lower bound above 20 points at 5 in-support conditions on LA and 4 on ST-Codecfake-OOD, so the effect is not confined to Transformer self-supervised representations. One additional encoder is descriptive evidence, not evidence of architecture-general behaviour, and the clean gate behind it is not weighting-robust: ECAPA-TDNN passed the registered five-seed gate on both corpora, but under equal weighting of the three unique split configurations (Section~\ref{sec:III}) ST-Codecfake-OOD moves from PASS to FAIL, 0.8681 against 0.8366 either side of the 0.85 gate, while LA remains clean-qualified. \textbf{ECAPA therefore provides supporting cross-representation generality evidence, robustly on LA and weighting-sensitive on ST-Codecfake-OOD.} WavLM-Base+ and W2V2-BERT 2.0 clear the gate on both corpora under both weightings (lowest value 0.8930), so the registered decision above is unaffected.

The Proxy-Anchor head shows the failure is not an artifact of a weak linear boundary. It qualifies against its clean gate and is non-inferior to the linear head on the same representation (PA -- LR = $-$0.14 points, lower bound $-$0.99, on LA; $-$1.72, lower bound $-$2.78, on ST-Codecfake-OOD, against a $-$3-point margin), yet degrades comparably under stress: +41.7 [+37.7, +45.6] points at EnCodec-6k on LA and +41.0 [+30.1, +51.9] on ST-Codecfake-OOD. The stress figures are weighting-robust, keeping a lower bound above 10 points on both corpora under equal weighting (+41.1 [+36.8, +45.3] and +41.5 [+30.4, +52.5]). The ST clean non-inferiority qualification is not: its lower bound moves from $-$2.78 to $-$4.28, crossing the $-$3-point margin, while LA stays non-inferior at $-$1.19. \textbf{The registered Proxy-Anchor instantiation also exhibits large codec-stress degradation on both corpora; however, its ST clean non-inferiority qualification relative to LR is sensitive to duplicate-split weighting.} The unqualified head comparison therefore holds for LA only, and a stronger metric-learning objective does not rescue attribution under transport.

On each corpus the single largest loss falls at a \textbf{conventional} operating point --- Opus at 6 kbit/s on LA and MP3 at 8 kbit/s on ST-Codecfake-OOD --- so this is not a neural-codec-specific phenomenon.

\section{Result 2: Degradation Is Strongly Condition- and Representation-Dependent}\label{sec:VI}
\begin{figure*}[t]
\centering
\includegraphics[width=\textwidth]{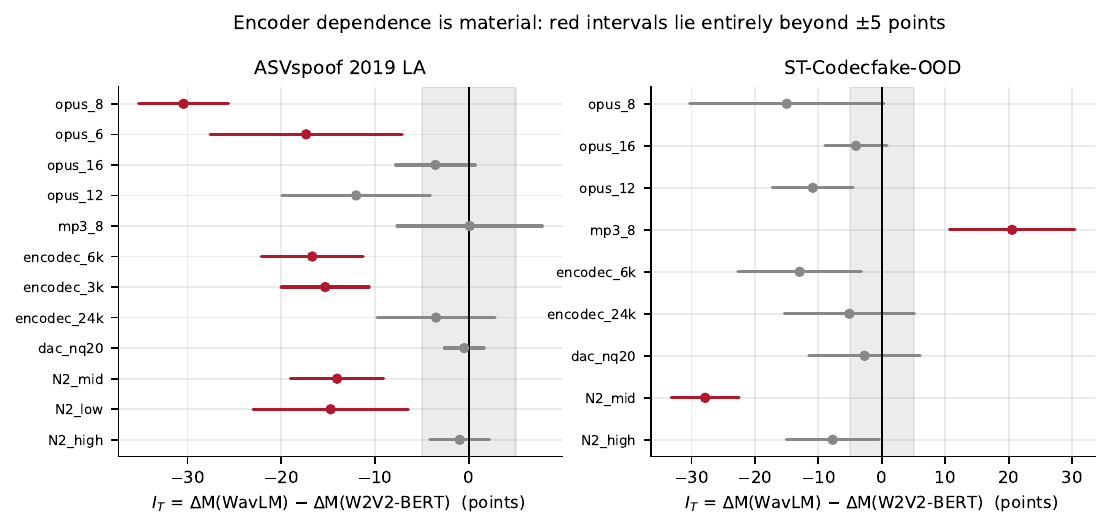}
\caption{Encoder interaction $I_T=\Delta$M(WavLM)$-\Delta$M(W2V2-BERT) with simultaneous bands. Red intervals lie entirely beyond the prespecified $\pm$5-point materiality margin.}
\label{fig:3}
\end{figure*}

A single stress number summarises neither of the two directions an engineer would want to average over. Within the same frozen in-support grid, losses indistinguishable from zero sit beside losses above 50 points: for WavLM-Base+ on LA the range runs from $-$0.4 [$-$2.4, +1.7] points at DAC $n_q$ = 20 --- an interval covering zero --- to +53.5 [+43.5, +63.6] at MP3-8, with DAC-high and Opus-16 near +11 and EnCodec-3k, DAC-low and Opus-6 above +39; on ST-Codecfake-OOD, from +2.6 [$-$1.0, +6.2] at Opus-16 to +70.3 [+63.0, +77.5] at MP3-8. Table~\ref{tab:6} and Table~\ref{tab:7} give every condition. Which of these a deployment meets is a property of the operating point, not of the attribution system.

Degradation is not a property of the condition alone either. Fig.~\ref{fig:3} reports the encoder interaction $I_T$ = $\Delta$M(WavLM) -- $\Delta$M(W2V2-BERT) with simultaneous bands, against a prespecified $\pm$5-point margin of practical materiality. On LA, six of twelve in-support conditions have intervals lying entirely beyond $\pm$5 points: N2\_low ($-$14.72 [$-$23.00, $-$6.45]), N2\_mid ($-$14.04 [$-$19.01, $-$9.08]), EnCodec-3k ($-$15.31 [$-$19.99, $-$10.63]), EnCodec-6k ($-$16.69 [$-$22.12, $-$11.26]), Opus-6 ($-$17.34 [$-$27.59, $-$7.10]) and Opus-8 ($-$30.43 [$-$35.21, $-$25.64]). On ST-Codecfake-OOD, two of nine do: N2\_mid ($-$27.82 [$-$33.16, $-$22.49]) and mp3\_8 (+20.53 [+10.66, +30.41]).

The sign of the last matters: on every LA condition clearing the margin, W2V2-BERT degrades more than WavLM, whereas at mp3\_8 on ST-Codecfake-OOD the ordering reverses. Across all three encoders the registered omnibus over pairwise interactions at in-support conditions \textbf{rejects on both corpora}, with 11 and 7 contrasts beyond $\pm$5 points. Which representation is more robust is not stable across two corpora, and a robustness number measured on one does not transfer to another.

Equal weighting of the three unique split configurations leaves the LA set of material conditions unchanged and raises the ST-Codecfake-OOD count from two to three, Opus-12 joining as its interaction moves further from zero ($-$10.87 [$-$17.23, $-$4.52] registered, $-$13.86 [$-$20.49, $-$7.23] equally weighted). Every retained interaction keeps its sign.

\section{Result 3: The Matched Comparison Was Not Estimable, and Fidelity Measures Rank Conditions Differently}\label{sec:VII}
\begin{figure*}[t]
\centering
\includegraphics[width=\textwidth]{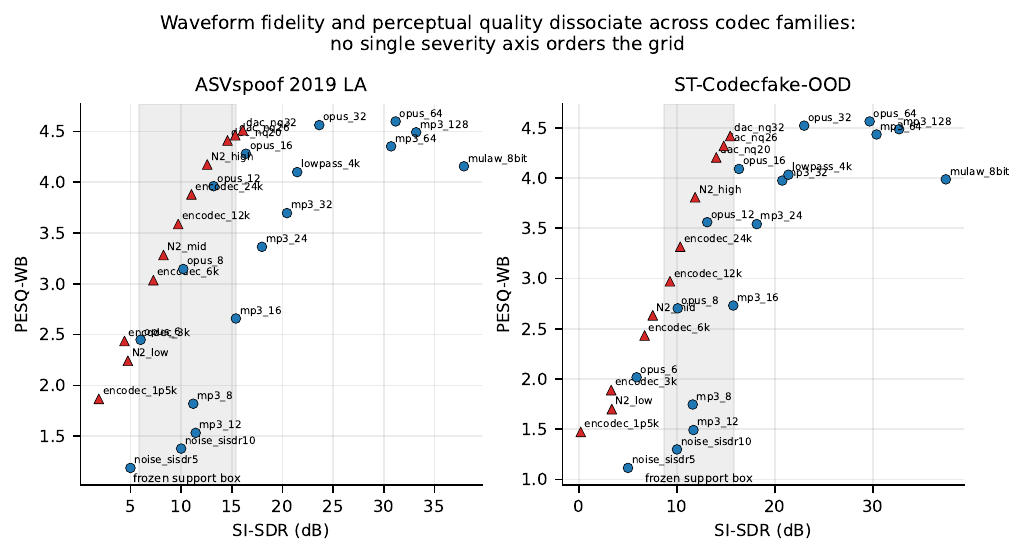}
\caption{Waveform fidelity (SI-SDR) against perceptual quality (PESQ-WB) for every condition in the frozen grid. Neural codecs (triangles) sit above conventional codecs (circles) at equal SI-SDR. The shaded band is the frozen support box.}
\label{fig:2}
\end{figure*}

\begin{table*}[t]
\caption{Non-codec controls: mean residual from the fidelity-only model M0 fitted on codec points, with simultaneous bands.}
\label{tab:3}
\centering\small
{\hyphenpenalty=9000\exhyphenpenalty=9000
\begin{tabular}{@{}>{\raggedright\arraybackslash}p{0.123\linewidth}>{\raggedright\arraybackslash}p{0.207\linewidth}>{\raggedright\arraybackslash}p{0.148\linewidth}>{\raggedright\arraybackslash}p{0.410\linewidth}@{}}
\toprule
control & mean residual from M0 (pts) & simultaneous band & verdict \\
\midrule
lowpass\_4k & +14.8 & [+6.2, +23.4] & more degradation than the fidelity-only account predicts \\
mulaw\_8bit & +13.6 & [-5.8, +33.0] & inconclusive \\
noise\_sisdr10 & -3.4 & [-12.0, +5.3] & inconclusive \\
noise\_sisdr5 & -19.5 & [-30.6, -8.5] & less degradation than predicted \\
\bottomrule
\end{tabular}}
\end{table*}

No single measured fidelity axis orders the codec grid for this purpose.

\textbf{The registered matching design lacked common support.} The four-dimensional rule of Section~\ref{sec:III} admitted exactly one pair per corpus and failed its gate. Fig.~\ref{fig:2} shows why: at equal SI-SDR, neural codecs score \textbf{0.4 to 2.7 PESQ-WB points higher} than MP3 and Opus across the frozen grid. Waveform fidelity and perceptual quality do not co-vary across codec families, and the matched contrast is therefore \textbf{not estimable under the preregistered matching rule and tested grid}. The same dissociation produced the empty support box on our sensitivity corpus. The verdict is scoped as in Section~\ref{sec:III}. A study that assumes such matching is available should verify common support before relying on it.

\textbf{The two fidelity measures rank the conditions differently.} Ordering the in-support conditions by mean SI-SDR and then by mean PESQ-WB does not give the same list. The clearest case is also the practically important one: MP3 at 8 kbit/s ranks \textbf{5th of 12} on SI-SDR on LA (11.49 dB) and \textbf{last} on PESQ-WB (1.85), and \textbf{4th of 9} on SI-SDR on ST-Codecfake-OOD (11.93 dB) and again \textbf{last} on PESQ-WB (1.81) --- while being the condition that costs the most attribution performance on both corpora. EnCodec-3k inverts the same disagreement on LA, ranking last on SI-SDR (4.65 dB) but 9th of 12 on PESQ-WB. An engineer selecting a "worst case" operating point from SI-SDR alone would not select the condition that actually does the most damage.

\textbf{Non-codec controls break the fidelity-only account.} If degradation were a function of measured fidelity alone, the residual of a non-codec distortion from a fidelity-only model fitted on codec points would be near zero. Table~\ref{tab:3} shows it is not: a 4 kHz bandwidth bottleneck degrades attribution \textbf{more} than that account predicts (+14.8 [+6.2, +23.4] points) and 5 dB additive noise \textbf{less} ($-$19.5 [$-$30.6, $-$8.5]). The \textmu{}-law control (+13.6 [$-$5.8, +33.0]) and 10 dB noise ($-$3.4 [$-$12.0, +5.3]) are inconclusive, and \textmu{}-law sits outside the codec support box, so its residual involves extrapolation. Only bandwidth and 5 dB noise are robust, and they point in opposite directions.

What survives depends on the \textit{structure} of the distortion, not only its magnitude, which is more than a single "codec severity" covariate can represent.

\section{Secondary Analysis: Conditional Association within the Support Region}\label{sec:VIII}
\begin{figure*}[t]
\centering
\includegraphics[width=\textwidth]{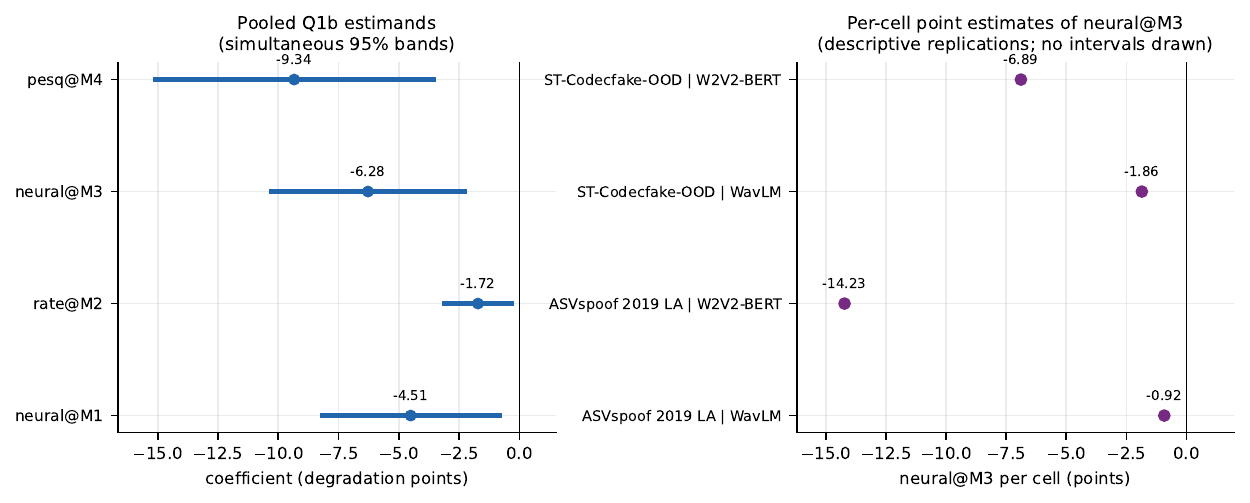}
\caption{Left: pooled Q1b estimands with simultaneous 95\% bands. Right: per-cell point estimates of the neural coefficient at M3 (descriptive replications); no per-cell intervals are drawn.}
\label{fig:4}
\end{figure*}

This analysis is \textbf{secondary and exploratory}: it replaced a comparison that was not estimable (Section~\ref{sec:III}), its support rule is calibration-informed and post-registration, and none of the three contributions rests on it. It was registered in advance, and its null results are reported in full.

Within the frozen support region we fit a nested ladder of least-squares models on a signed degradation indicator, with corpus and encoder indicators throughout, and test four coefficients under simultaneous bands against a registered practical margin of $\delta_D$ = 5 points (Table~\ref{tab:2}, Fig.~\ref{fig:4}). Appendix B states the response, the five models M0--M4 and the six estimands exactly.

\begin{table}[t]
\caption{Q1b estimands with simultaneous 95\% bands, and the registered truth table.}
\label{tab:2}
\centering\small
{\hyphenpenalty=9000\exhyphenpenalty=9000
\begin{tabular}{@{}lll@{}}
\toprule
estimand & point & simultaneous 95 \% band \\
\midrule
neural@M1 & $-$4.51 & [$-$8.20, $-$0.82] \\
rate@M2 & $-$1.72 & [$-$3.12, $-$0.32] \\
neural@M3 & $-$6.28 & [$-$10.30, $-$2.26] \\
PESQ@M4 & $-$9.34 & [$-$15.12, $-$3.56] \\
$\Delta R^2$ (rate -- family) & +0.0002 & [$-$0.0052, +0.0042] \\
$\Delta R^2$ (PESQ) & +0.0110 & [+0.0034, +0.0230] \\
\bottomrule
\end{tabular}}
\vspace{3pt}\par\footnotesize Registered decisions: H1 inconclusive; H2 inconclusive; H3 inconclusive; H4 supported. Negative coefficient = associated with less degradation, holding the others fixed.
\end{table}

The registered truth table returns \textbf{H4 supported; H1, H2 and H3 inconclusive.} Three coefficients have intervals excluding zero, yet none satisfies the stronger registered rule its hypothesis requires. A non-zero coefficient is not a decision.

\textbf{H4 --- perceptual quality adds information beyond waveform fidelity, intelligibility, rate and family.} PESQ@M4 excludes zero and the R$^2$ increment has a lower bound above zero, so H4 is supported under its registered rule. The effect is \textbf{substantively modest} --- roughly one percentage point of additional explained variance --- and it is non-redundancy evidence inside one support region, not a causal statement about why attribution fails; practically it adds nothing to Section~\ref{sec:VII}.

\textbf{Two sensitivity analyses leave the truth table intact.} The registered primary bootstrap resamples components \textit{without} stratification (Section~\ref{sec:III}); a class-signature-stratified variant, run as a sensitivity audit, returns PESQ@M4 $-$9.34 [$-$14.02, $-$4.67] and $\Delta R^2$ 0.0110 [0.0043, 0.0193] with unchanged H1--H4 decisions, and H4 also survives a quadratic functional form. Under equal weighting of unique split configurations the rate@M2 band no longer excludes zero, again without changing the truth table: H2 was inconclusive under the registered rule and remains so under both weightings.

\textbf{The pooled neural coefficient is negative but not uniform.} Neural conditions are associated with \textit{less} degradation after adjustment. Across the four corpus $\times$ encoder cells the per-cell point estimates are $-$14.23 for LA | W2V2-BERT, $-$0.92 for LA | WavLM, $-$1.86 for ST | WavLM and $-$6.89 for ST | W2V2-BERT. Only the LA | W2V2-BERT interval excludes zero; the other three include zero. The pooled estimate is numerically dominated by one cell, and collinearity constrains its interpretation on LA (PESQ variance-inflation factor 7.5, PESQ--STOI correlation 0.88). We report it as a conditional association only.

\section{Operational Diagnostic: The Registered Overwrite Rule Was Not Confirmed}\label{sec:IX}
\begin{figure*}[t]
\centering
\includegraphics[width=\textwidth]{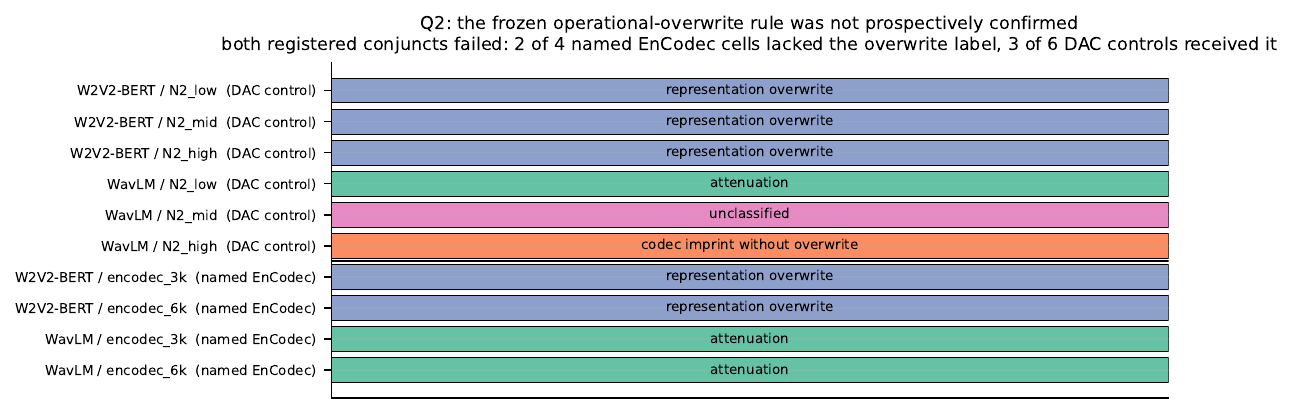}
\caption{Labels assigned by the frozen overwrite rule to the ten registered LA cells. Both registered conjuncts failed; descriptively, the labels separate by representation rather than by codec family.}
\label{fig:5}
\end{figure*}

\subsection{The rule and what it was for}

Our development analysis suggested a mechanistic reading of neural-codec degradation: under selected EnCodec conditions the codec appeared not merely to attenuate the generator signature but to replace it. The reading is attractive --- compatible with how neural codecs are built, and convertible into simple operational guidance --- so rather than report it we froze it as an executable decision procedure, its numerical core verified byte-identical to the development implementation before registration, and tested it prospectively. What follows is an \textbf{operational diagnostic}: whether such a rule, fixed in advance, flags the conditions under which attribution fails. It is not a mechanism study.

Each (encoder, condition) cell receives exactly one label from four interval conjuncts with margins fixed a priori at $\varepsilon$ = 0.05: \textbf{selectivity} of the codec direction, \textbf{margin loss} of the generator signal, \textbf{beyond shrinkage} relative to an isotropic-shrinkage null, and \textbf{restoration} when the codec mean shift is removed. A cell is labelled \textbf{representation overwrite} only if all four hold; weaker combinations receive one of four other labels. Appendix B states the statistics, the conjuncts and the labels exactly.

The registered decision was an intersection--union test fixed before execution: the account is confirmed only if \textbf{all four} named LA EnCodec cells receive the overwrite label \textbf{and none} of the six DAC control cells does. The named targets were chosen from development; the DAC cells are controls because a universal "neural codec $\Rightarrow$ overwrite" account predicts they behave like EnCodec, whereas an EnCodec-specific account predicts they do not.

\subsection{The registered outcome}

\textbf{Both conjuncts failed} (Fig.~\ref{fig:5}). Two of four named EnCodec cells did not receive the overwrite label, and three of six DAC control cells did. The registered decision is \textbf{C2 not prospectively confirmed}.

The duplicate-split issue of Section~\ref{sec:III} does not reach this decision: Q2 shuffles clips within each retained class --- and, in the grouped variant, the source clusters --- \textit{after} the class hold-out is drawn, so seeds that coincide on the held-out set still differ. All 72 seed-level conjunct evaluations were distinct, and no Q2 quantity is recomputed under the equal-weighting sensitivity.

\subsection{Descriptive inspection of the frozen labels}

The labels the frozen rule assigned are worth inspecting, \textbf{descriptively and post hoc}: encoder dependence of the label was not a registered hypothesis, and no significance test is attached to it.

Across the ten registered LA cells, WavLM-Base+ received the overwrite label in \textbf{0 of 5} cells and W2V2-BERT 2.0 in \textbf{5 of 5}. WavLM was labelled \textit{attenuation} at EnCodec-6k, EnCodec-3k and DAC-low, \textit{codec imprint without overwrite} at DAC-high and \textit{unclassified} at DAC-mid; W2V2-BERT received \textit{representation overwrite} at both EnCodec operating points and all three DAC operating points. By codec family the same ten cells split \textbf{2 of 4} named EnCodec and \textbf{3 of 6} DAC controls: the grouping the rule was built to separate does not separate them.

The supporting transport analysis on ST-Codecfake-OOD, which is outside the registered decision, reached no overwrite cell at all: every EnCodec and DAC cell, for either encoder, was labelled \textit{selective codec imprint with generator-margin loss}, \textit{codec imprint without overwrite} or \textit{unclassified}. One implementation limitation is disclosed: the frozen rule applies matched-class exclusion only to EnCodec-named conditions, so on ST-Codecfake-OOD the DAC provenance class is not excluded under DAC conditions. LA is unaffected, as no LA attack class is a codec.

\subsection{What the diagnostic shows}

The registered rule did not separate the codec families it was built to separate: it fired on three of six DAC control cells and missed two of four named EnCodec cells. \textbf{As an operational flag for codec-induced attribution failure, fixed in advance and applied prospectively, it did not work on this grid.}

The labels instead tracked the representation, consistent with the representation dependence of Section~\ref{sec:VI}. That pattern does not establish that overwrite never occurs, that neural codecs do not overwrite, or that encoder architecture causes any of it, and on our data the rule does not support the deployment guidance it implies.

\subsection{Why the rule looked promising on development data}

On a development corpus pair, the same four-conjunct procedure applied to ST-Codecfake with WavLM-Base+ gave exactly the pattern the account describes at all three EnCodec operating points --- a selectively decodable codec direction well clear of the $\varepsilon_C$ margin together with a large negative generator-margin change (EnCodec-6k C = +0.230 [+0.212, +0.247] with G = $-$0.334; EnCodec-3k +0.191 [+0.179, +0.203] with G = $-$0.409; EnCodec-1.5k +0.146 [+0.137, +0.154] with G = $-$0.466) --- while MP3 and Opus on the same encoder and corpus reached \textit{codec imprint without overwrite} with margin changes an order of magnitude smaller (|G| $\leq$ 0.031). The prospective targets of this section were selected from that observation.

Two weaknesses were recorded before any label was interpreted. The pre-registered permutation-null conjunct was uninformative by construction --- with four codec families, a label-exchange null for an own-minus-max-other statistic has a minimum attainable tail near 1/4! --- so it was waived; and the isotropic-shrinkage control was non-binding, the shrinkage statistic being indistinguishable from zero at every condition. The development labels are therefore labels under a modified rule, and were reported as such.

The prospective result contradicts the development reading at exactly the cells it was built from: WavLM at EnCodec-6k and EnCodec-3k, labelled \textit{representation overwrite} in development, was labelled \textit{attenuation} prospectively. \textbf{Development evidence was hypothesis-generating and carries no confirmatory weight.}

\section{Discussion}\label{sec:X}

\textbf{Why "neural codec is worse" is too simple.} The largest single losses on both corpora occur at conventional operating points (Section~\ref{sec:V}), and after adjustment the neural-codec association points toward \textit{less} degradation and is carried by one of four cells (Section~\ref{sec:VIII}). Because the families occupy different regions of the fidelity space, a comparison that does not establish common support compares operating points rather than families. We make no causal family claim in either direction.

\textbf{Outcome and mechanism are different claims.} That codec transport costs a large amount of attribution performance holds within our tested scope: it is the registered primary decision, on both corpora, for the two deciding representations and two head types, and descriptively for a third. Why it happens is not something this study establishes; the one mechanism-shaped rule we froze and tested prospectively was not confirmed (Section~\ref{sec:IX}).

\textbf{What should benchmarks report?} Three recommendations follow from the results above.

\textbf{A. Report codec stress under stated calibration and threshold conditions; a clean accuracy figure alone is insufficient.} All three encoders exceed 0.85 clean Macro-F1 and two exceed 0.97, yet lose 53.5 and 70.3 points at a single in-grid operating point (Section~\ref{sec:V}). The gap between the clean number and the transported one is condition-specific and, on this evidence, large; because it also depends on where the threshold and calibration were set, one accuracy figure cannot carry it.

\textbf{B. Report several fidelity, perceptual and rate axes rather than one severity scalar, and declare lack of common support when a matched comparison is not estimable.} Our registered matching design failed its gate because the two families dissociate on waveform versus perceptual quality; the two measures rank the conditions differently, with the most damaging condition mid-grid on one and last on the other; and the non-codec controls depart from a fidelity-only account in opposite directions (Section~\ref{sec:VII}). A matched-fidelity comparison whose common support was never checked may not be estimable at all.

\textbf{C. Evaluate at least two materially different representations, or state conclusions as representation-bounded.} Encoder interaction exceeded the $\pm$5-point materiality margin at six of twelve in-support conditions on one corpus, the three-encoder omnibus rejected on both, and which representation is more robust reverses between corpora (Section~\ref{sec:VI}); descriptively, the diagnostic labels of Section~\ref{sec:IX} separated the representations rather than the codec families. On this evidence a robustness number measured on one encoder does not transfer to another.

Whether the degradation measured here is recoverable is an open question our evidence does not address, as are transcoding cascades --- our grid is single-stage throughout --- and the calibration behaviour of attribution scores under transport.

\section{Limitations}\label{sec:XI}

\begin{enumerate}
\item \textbf{Effective sample size.} 45 independent components on LA and 28 on ST-Codecfake-OOD. All intervals are component-level; clip counts are far larger and would overstate precision.
\item \textbf{The matched-fidelity comparison is not estimable under the preregistered matching rule and tested grid}, so the registered clauses that depended on it --- including the formal generality clause --- are non-executable. Where a matched conventional point was required, the nearest in-support conventional condition is substituted and labelled descriptive.
\item \textbf{Q1b is secondary, calibration-informed and post-registration.} Its support rule was selected on calibration data after a draft rule failed there, so residual design-fitting risk is acknowledged. Within it, H1, H2 and H3 are inconclusive rather than null --- non-zero coefficients that fail the registered practical-margin and incremental-fit rules do not become findings --- and H4 is modest: roughly one percentage point of additional explained variance, with PESQ collinearity on LA, and non-redundancy evidence only.
\item \textbf{Two corpora, three representations, one single-stage codec grid, closed-set tasks.} Every claim in this paper is bounded to that scope. No field-wide, open-set, cascade or architecture-general result is claimed, and the two corpora carry different label semantics, so attribution magnitudes are not compared across them.
\item \textbf{A disclosed Q2 implementation limitation} affects matched-class exclusion on ST-Codecfake-OOD under DAC conditions; LA is unaffected.
\item \textbf{Seed collisions and weighting sensitivity.} The five seeded runs realise three unique class-hold-out configurations per corpus (Section~\ref{sec:III}). Weighting those three equally leaves every primary decision unchanged, but moves two qualification outcomes on ST-Codecfake-OOD --- the ECAPA-TDNN clean gate and the Proxy-Anchor clean non-inferiority (Section~\ref{sec:V}) --- and leaves the rate@M2 band no longer excluding zero, without altering the H1--H4 decisions (Section~\ref{sec:VIII}). Q2 is unaffected. A future registration should draw class hold-outs without replacement across seeds.
\item \textbf{No mitigation is proposed or evaluated.}
\end{enumerate}

\section{Conclusion}\label{sec:XII}

For the closed-set provenance tasks, corpora, representations and single-stage codec grid tested here, clean attribution accuracy does not guarantee robustness after codec transport. Under a protocol whose analysis region was fixed from fidelity metadata before any attribution model was trained, transport removed a large fraction of attribution performance on both corpora, for the two deciding representations and two head types and descriptively for a third.

The loss is not summarised by one number. Inside the same in-support grid, conditions whose effect is indistinguishable from zero sit beside conditions costing more than 50 Macro-F1 points, and the two deciding representations differ beyond a prespecified materiality margin at half the LA conditions, with their ordering reversing between corpora. Nor is severity summarised by one fidelity axis: the registered matched-fidelity comparison was not estimable on this grid, the waveform and perceptual measures rank the conditions differently, and non-codec controls depart from a fidelity-only account in opposite directions.

A rule that would have flagged codec-induced failure in advance, frozen from development evidence and tested prospectively, was not confirmed. What was measured, rather than why it happens, is what this paper reports: robustness after transport has to be measured on the conditions, representations and operating points a deployment will actually see.

\section{Data and Code Availability}

Code and evaluation artifacts --- the registered protocol, the frozen support masks with their hashes, the figure-generation scripts, per-cell intervals, complete Q2 conjunct labels, and the correction history --- will be released in a future revision. The corpora are ASVspoof 2019 LA and ST-Codecfake-OOD, obtained from their original distributors under their respective licences; MLAAD is used only for the designated sensitivity arm. All prospective computation totalled 9.34 GPU-hours on a single RTX 4060 Ti (16 GB), run sequentially, with peak VRAM 3.4 GB; 57,057 clip-condition transformation operations completed with zero failures.

\section{Declaration on the Use of Generative AI}

Generative AI tools were used solely to assist with language editing, phrasing, and manuscript preparation. They were not used to formulate the research questions, generate scientific ideas, develop the methodology, design or conduct experiments, analyze results, or draw scientific conclusions. All scientific contributions, experimental decisions, analyses, and interpretations were developed and verified by the author, who takes full responsibility for the content of this manuscript.

\section*{Appendix A: Condition-Level Results}

\begin{table*}[!t]
\caption{Condition-level $\Delta$Macro-F1 with simultaneous bands, ASVspoof 2019 LA. Positive means attribution loss. Bands are calibrated over the registered in-support family only; out-of-support conditions carry a point estimate and a dash, and no decision weight.}
\label{tab:6}
\centering\footnotesize\renewcommand{\arraystretch}{0.88}
{\hyphenpenalty=9000\exhyphenpenalty=9000
\begin{tabular}{@{}>{\raggedright\arraybackslash}p{0.123\linewidth}>{\raggedright\arraybackslash}p{0.090\linewidth}>{\raggedright\arraybackslash}p{0.161\linewidth}>{\raggedright\arraybackslash}p{0.156\linewidth}>{\raggedright\arraybackslash}p{0.159\linewidth}>{\raggedright\arraybackslash}p{0.163\linewidth}@{}}
\toprule
condition & in support & WavLM-Base+ & W2V2-BERT 2.0 & ECAPA-TDNN & Proxy-Anchor \\
\midrule
N2\_high & yes & +11.6 [+8.7, +14.5] & +12.5 [+7.4, +17.7] & +10.8 [-1.8, +23.3] & +11.8 [-1.1, +24.7] \\
N2\_low & yes & +43.2 [+39.8, +46.6] & +58.0 [+50.7, +65.2] & +37.3 [+25.8, +48.9] & +63.0 [+60.7, +65.3] \\
N2\_mid & yes & +22.9 [+17.6, +28.1] & +36.9 [+29.0, +44.8] & +23.1 [+7.9, +38.3] & +42.2 [+38.2, +46.2] \\
dac\_nq20 & yes & -0.4 [-2.4, +1.7] & +0.1 [-2.2, +2.3] & +8.2 [-3.5, +19.9] & +0.1 [-1.1, +1.2] \\
encodec\_24k & yes & +13.9 [+9.2, +18.6] & +17.3 [+10.1, +24.6] & +19.8 [+6.7, +32.9] & +15.1 [+8.8, +21.4] \\
encodec\_3k & yes & +39.3 [+32.6, +46.0] & +54.6 [+48.6, +60.6] & +39.6 [+28.0, +51.1] & +54.0 [+49.4, +58.5] \\
encodec\_6k & yes & +26.7 [+20.5, +32.9] & +43.4 [+36.6, +50.2] & +28.5 [+13.8, +43.1] & +41.7 [+37.7, +45.6] \\
mp3\_8 & yes & +53.5 [+43.5, +63.6] & +53.4 [+43.9, +63.0] & +61.5 [+47.0, +75.9] & +48.0 [+41.9, +54.1] \\
opus\_12 & yes & +12.7 [+9.3, +16.2] & +24.7 [+16.3, +33.2] & +6.0 [-7.3, +19.3] & +22.4 [+19.0, +25.8] \\
opus\_16 & yes & +11.2 [+9.8, +12.5] & +14.7 [+9.6, +19.8] & +2.8 [-9.9, +15.5] & +11.7 [+8.9, +14.4] \\
opus\_6 & yes & +43.6 [+34.7, +52.6] & +61.0 [+56.8, +65.1] & +76.0 [+62.3, +89.6] & +54.0 [+51.0, +56.9] \\
opus\_8 & yes & +20.6 [+14.3, +26.9] & +51.0 [+42.8, +59.2] & +73.7 [+60.4, +87.1] & +41.7 [+37.3, +46.1] \\
encodec\_1p5k & no & +61.7 -- & +67.2 -- & +48.4 -- & +65.2 -- \\
lowpass\_4k & no & +2.3 -- & +19.3 -- & +76.7 -- & +12.3 -- \\
mp3\_128 & no & +0.5 -- & +1.9 -- & +11.5 -- & +0.7 -- \\
mp3\_32 & no & +11.3 -- & +17.5 -- & +31.2 -- & +15.9 -- \\
mp3\_64 & no & +0.1 -- & +1.9 -- & +9.7 -- & +0.5 -- \\
mulaw\_8bit & no & +0.6 -- & -0.1 -- & +0.9 -- & +0.8 -- \\
noise\_sisdr10 & no & +27.2 -- & +44.6 -- & +66.6 -- & +40.7 -- \\
noise\_sisdr5 & no & +31.8 -- & +65.4 -- & +72.2 -- & +63.8 -- \\
opus\_32 & no & +0.7 -- & -0.5 -- & +0.7 -- & -0.5 -- \\
opus\_64 & no & +0.0 -- & +0.0 -- & +0.2 -- & -0.2 -- \\
\bottomrule
\end{tabular}}
\end{table*}

\begin{table*}[!t]
\caption{Condition-level $\Delta$Macro-F1 with simultaneous bands, ST-Codecfake-OOD. Conventions as in Table~\ref{tab:6}.}
\label{tab:7}
\centering\footnotesize\renewcommand{\arraystretch}{0.88}
{\hyphenpenalty=9000\exhyphenpenalty=9000
\begin{tabular}{@{}>{\raggedright\arraybackslash}p{0.123\linewidth}>{\raggedright\arraybackslash}p{0.090\linewidth}>{\raggedright\arraybackslash}p{0.161\linewidth}>{\raggedright\arraybackslash}p{0.156\linewidth}>{\raggedright\arraybackslash}p{0.159\linewidth}>{\raggedright\arraybackslash}p{0.163\linewidth}@{}}
\toprule
condition & in support & WavLM-Base+ & W2V2-BERT 2.0 & ECAPA-TDNN & Proxy-Anchor \\
\midrule
N2\_high & yes & +6.6 [+1.0, +12.2] & +14.4 [+2.6, +26.1] & +21.6 [+14.1, +29.1] & +10.9 [+6.2, +15.5] \\
N2\_mid & yes & +13.4 [+6.5, +20.2] & +41.2 [+34.5, +47.8] & +39.2 [+30.5, +47.9] & +42.9 [+34.7, +51.1] \\
dac\_nq20 & yes & +3.2 [+0.5, +5.9] & +5.9 [-4.6, +16.5] & +13.0 [+5.3, +20.6] & +3.9 [+1.0, +6.7] \\
encodec\_24k & yes & +12.6 [+5.7, +19.6] & +17.8 [+6.0, +29.5] & +29.7 [+19.5, +39.9] & +17.8 [+10.1, +25.4] \\
encodec\_6k & yes & +22.2 [+14.5, +29.9] & +35.1 [+21.2, +49.1] & +37.6 [+26.2, +48.9] & +41.0 [+30.1, +51.9] \\
mp3\_8 & yes & +70.3 [+63.0, +77.5] & +49.8 [+41.6, +57.9] & +75.2 [+66.5, +84.0] & +54.2 [+45.8, +62.6] \\
opus\_12 & yes & +5.1 [-0.4, +10.5] & +16.0 [+9.2, +22.7] & +18.2 [+12.4, +24.0] & +17.0 [+8.8, +25.2] \\
opus\_16 & yes & +2.6 [-1.0, +6.2] & +6.6 [+2.5, +10.8] & +12.1 [+5.9, +18.2] & +6.7 [+1.9, +11.5] \\
opus\_8 & yes & +21.0 [+7.1, +34.8] & +35.9 [+19.9, +52.0] & +68.3 [+59.2, +77.5] & +30.5 [+20.9, +40.2] \\
N2\_low & no & +48.3 -- & +63.9 -- & +51.1 -- & +70.5 -- \\
encodec\_1p5k & no & +62.2 -- & +57.8 -- & +50.8 -- & +74.0 -- \\
encodec\_3k & no & +41.3 -- & +44.1 -- & +47.4 -- & +55.9 -- \\
lowpass\_4k & no & +3.8 -- & +18.3 -- & +61.7 -- & +9.4 -- \\
mp3\_128 & no & +0.9 -- & +1.9 -- & +9.3 -- & +1.6 -- \\
mp3\_32 & no & +7.9 -- & +27.2 -- & +26.9 -- & +22.0 -- \\
mp3\_64 & no & +1.9 -- & +3.6 -- & +10.4 -- & +3.6 -- \\
mulaw\_8bit & no & +1.5 -- & +3.5 -- & +6.0 -- & +6.6 -- \\
noise\_sisdr10 & no & +56.8 -- & +52.8 -- & +67.4 -- & +71.3 -- \\
noise\_sisdr5 & no & +65.7 -- & +70.8 -- & +72.5 -- & +80.1 -- \\
opus\_32 & no & +0.6 -- & +1.5 -- & +4.8 -- & +2.0 -- \\
opus\_64 & no & +0.5 -- & +1.0 -- & -1.3 -- & +0.5 -- \\
\bottomrule
\end{tabular}}
\end{table*}

Table~\ref{tab:6} and Table~\ref{tab:7} report every condition in the frozen grid for all four models on each corpus. Simultaneous bands exist only for in-support conditions, over which the max-|t| calibration was computed; out-of-support point estimates carry no intervals and no decision weight.

\section*{Appendix B: Analysis Specifications}

\textbf{B.1 The Q1b model ladder (Section~\ref{sec:VIII}).} For clip $i$, let $a_i$ indicate that its clean prediction is correct and $b_{ic}$ that its prediction under condition $c$ is correct. The signed degradation response is

\begin{equation*}
y_{ic} = 100\,(a_i - b_{ic}) \in \{-100,\, 0,\, +100\},
\end{equation*}

so positive means the condition destroyed a correct clean decision and negative means it repaired an incorrect one. Write $s_{ic}$ for SI-SDR, $t_{ic}$ for STOI, $q_{ic}$ for PESQ-WB, $r_c = \log_2(\text{kbit/s})$ for the nominal rate, and $f_c = \mathbf{1}[c\ \text{is a neural codec}]$ for the family indicator, and let $Z_i$ collect the corpus and encoder indicators (one dropped level each). The five models are

\begin{equation*}
\begin{aligned}
\text{M0:}\quad & y = \beta_0 + \beta_s s + \beta_t t + Z\gamma \\
\text{M1:}\quad & y = \beta_0 + \beta_s s + \beta_t t + \beta_f f + Z\gamma \\
\text{M2:}\quad & y = \beta_0 + \beta_s s + \beta_t t + \beta_r r + Z\gamma \\
\text{M3:}\quad & y = \beta_0 + \beta_s s + \beta_t t + \beta_r r + \beta_f f + Z\gamma \\
\text{M4:}\quad & y = \beta_0 + \beta_s s + \beta_t t + \beta_r r + \beta_f f + \beta_q q + Z\gamma
\end{aligned}
\end{equation*}

each fitted by ordinary least squares. The four tested estimands are neural@M1 $= \beta_f$ in M1, rate@M2 $= \beta_r$ in M2, neural@M3 $= \beta_f$ in M3 and PESQ@M4 $= \beta_q$ in M4; the two incremental-fit terms are $\Delta R^2(\text{rate} - \text{family}) = (R^2_{M2} - R^2_{M0}) - (R^2_{M1} - R^2_{M0})$ and $\Delta R^2(\text{PESQ}) = R^2_{M4} - R^2_{M3}$. Each model is refitted within every split seed and bootstrap replicate; each estimand is the mean of its per-seed fits.

\textbf{B.2 The frozen overwrite rule (Section~\ref{sec:IX}).} For an (encoder, condition) cell let $Z_c$, $Z_t$ be the clean and transformed embeddings of the same clips with means $\bar z_c$, $\bar z_t$; $m(\cdot)$ the per-clip generator margin, own-class similarity minus the best competing class; $d_f$ the unit mean-shift direction induced by codec family $f$, i.e. the normalised difference between the transformed and clean means under $f$; $\hat{\delta}$ the per-clip normalised clean-to-transformed shift; and an overbar a mean over the cell's clips. Define

\begin{equation*}
\begin{aligned}
S_f &= \overline{\langle \hat{\delta},\, d_f \rangle}, \qquad C = S_{\text{own}} - \max_{f \neq \text{own}} S_f, \\
G &= \overline{m(Z_t) - m(Z_c)}, \\
K &= \overline{\lVert Z_t - \bar z_t \rVert^2} \big/ \overline{\lVert Z_c - \bar z_c \rVert^2}, \\
Z_{\text{sh}} &= \bar z_c + \sqrt{K}\,(Z_c - \bar z_c), \\
G_{\text{sh}} &= \overline{m(Z_{\text{sh}}) - m(Z_c)}, \\
G_{\text{res}} &= \overline{m(Z_t - d_{\text{own}}) - m(Z_t)}.
\end{aligned}
\end{equation*}

With margins fixed a priori at $\varepsilon_C = \varepsilon_G = 0.05$ and percentile intervals from the cluster bootstrap, the four conjuncts are

\begin{itemize}
\item \textbf{selectivity}  $\mathrm{lo}(C) > \varepsilon_C$ --- the cell moves along its own codec family's direction more than any other's;
\item \textbf{margin loss}  $\mathrm{hi}(G) < -\varepsilon_G$ --- the generator margin falls;
\item \textbf{beyond shrinkage}  $\mathrm{hi}(G - G_{\text{sh}}) < 0$ --- the fall exceeds what isotropic shrinkage by the observed factor $\sqrt{K}$ would produce;
\item \textbf{restoration}  $\mathrm{lo}(G_{\text{res}}) > 0$ --- removing the codec mean shift recovers margin.
\end{itemize}

A cell is labelled \textbf{representation overwrite} only if all four hold. Otherwise it is \textit{selective codec imprint with generator-margin loss} (selectivity and margin loss), \textit{codec imprint without overwrite} (selectivity with the margin retained, i.e. $\mathrm{lo}(G) > -\varepsilon_G$ and $\mathrm{hi}(G) < \varepsilon_G$), \textit{attenuation} (margin loss without selectivity), or \textit{unclassified}. Restoration removes the codec \textbf{mean shift} rather than projecting onto the complement of $d_{\text{own}}$; the two differ unless the cloud is isotropic about that direction.

\end{document}